\documentstyle[epsfig]{article}
\begin{document}
\pagestyle{plain}
\topmargin=-2mm
\textwidth=175mm
\textheight=230mm
\setcounter{page}{1}
\def\Journal#1#2#3#4{{#1} {\bf #2}, #3 (#4)}

\def\NCA{\em Nuovo Cimento}
\def\NIM{\em Nucl. Instrum. Methods}
\def\NIMA{{\em Nucl. Instrum. Methods} A}
\def\NPA{{\em Nucl. Phys.} A}
\def\NPB{{\em Nucl. Phys.} B}
\def\PLB{{\em Phys. Lett.}  B}
\def\PRL{\em Phys. Rev. Lett.}
\def\PRD{{\em Phys. Rev.} D}
\def\ZPC{{\em Z. Phys.} C}
\vspace{.5cm}
\begin{center}
{\Huge g-2 in composite models of leptons}

\vspace{.2in}
Yuan-Ben Dai, Chao-Shang Huang, and Ailin Zhang

\vspace{.2in}
Institute of theoretical Physics, P. O. Box 2735, Beijing, 100080,
P. R. China
\end{center}

\begin{center}
\begin{minipage}{5in}
\baselineskip=0.25in
\begin{center} ABSTRACT\end{center}
Based on the bound state description of the muon and general
relativistic covariant quantum field theory, we illustrate with a
simple composite model that the observed deviation of
$(g-2)_{\mu}$ can be a demonstration of the substructure of the muon
and give the constraints on the radius of the muon in different cases
of  light constituents and heavy constituents.
\end{minipage}
\end{center}
\vfill



\newpage
\section{Introduction}
\indent
Recently, the E821 experiment at the Brookhaven National Laboratory \cite{brown} announced
their latest measurement of the anomalous magnetic moment of the muon
\begin{eqnarray}
a_\mu={g-2\over 2}=11659202(14)(6)\times 10^{-10}.
\end{eqnarray}
In the Standard Model(SM), the anomalous magnetic moment, $a_\mu^{SM}=a_\mu ^{QED}+a_\mu ^{Had}+a_\mu ^{EW}$, is estimated to
be\cite{cm}

\begin{eqnarray}
a_\mu =116591597(67)\times 10^{-11},
\end{eqnarray}
where the error is mainly from $a_\mu ^{Had}$. An estimate of
hadronic contributions has been renewed recently \cite{lo}. From
these results one obtains present deviation from SM
\begin{eqnarray}\label{da}
\Delta a_\mu =a_\mu^{exp}-a_\mu^{SM}=(4.3\pm 1.6)\times 10^{-9}.
\end{eqnarray}
The size of the deviation is hotly debated due to the uncertainties in the hadron vacuum
polarization in $(g-2)_{\mu}$~\cite{nath}.
The deviation may give a hint of new physics. It is well-known
that the progress to investigate physics beyond SM, new physics,
is essentially along two different directions. One is to extend
the symmetry of SM, such as supersymmetry, the fourth generation,
more Higgs doublets, grand unification etc. or to assume the
presence of extra dimensions, while keeping the assumption that
all the SM particles as well as new particles are fundamental. The
other is to assume the substructure of SM particles, such as
composite models of lepton, quark and weak gauge boson,
technicolor models, etc. Inspired by the exciting $2.6\sigma$
excess, people have suggested many possible explanations of the
deviation, mostly along the first direction~\cite{fun,many}. In
particular, supersymmetric loop effects with $m_{SUSY}\simeq
55\sqrt{\tan\beta}$ Gev and the radiative mass mechanism at $1-2$ TeV
scale have been reviewed in \cite{cm,nath}.  The reason of the
deviation has also been probed, along the second direction, in
technicolor models~\cite{yan} and preonic models~\cite{drr,lane}. In
ref.\cite{drr}, the deviation is attributed to the presence of
heavy exotic colored lepton and extra Z-boson states arising in a
preonic model and essentially the bound state description of a
lepton is not touched in their calculations. In ref.\cite{lane}
the model-independent limit on muon substructure has been given by
using the previous formula given in ref.\cite{bd}.  The constraint from
the anomalous magnetic moments of leptons on possible substructure
of leptons has been analyzed in terms of a general formalism for
describing a bound state in early 80's\cite{dh,bd}.
 In the non-relativistic theory of a bound state , where the
binding energy and the inverse radius $R^{-1}$ are much smaller
than the mass of the bound state , the magnetic moment of the
bound state is the vectorial sum of the magnetic moments of the
constituents. This would contradicts the precise experimental data
for $a_\mu $ if $\mu \;$is such a bound state ~\cite{lipkin}. It
has been shown that this is not the case for the relativistic
bound state~\cite{bd,dh}. In this paper, we will consider the
theoretical implication of the E821 new result in a composite
model of the lepton, based on the bound state description of the muon and
the general relativistic covariant quantum field theory. We
generalize the previous investigations to higher orders in
$\alpha$ and including both the heavy and light constituent cases.
We shall repeat some previous derivations in order to make the
paper self-contained. For the sake of simplicity we calculate the
anomalous magnetic moment of a muon in a simple composite model in
which the lepton is assumed to be a bound state composed of a
fermion and a scalar boson.
 We show that the deviation, $\delta a_{\mu}$, can be a
signal for compositeness of the muon and poses a constraint on the
charge radius of the muon (and the masses of constituents if
constituents are heavy).

Our paper is organized as follows. In Sec. II, we present the
general form of the Bethe-Salpeter (BS) bound state wave function
of a lepton, calculate the lowest order matrix element of the
electric current of a composite particle and estimate the order of
magnitude for the anomalous magnetic moment of a composite lepton at
the lowest order in $\alpha$. Using the spectrum representation of the
wave function and the general properties of the four-point Green function
containing the lepton bound state pole, the $O(\alpha)$ correction to the anomalous
magnetic moment is estimated in Sec. III. The last section is
devoted to conclusions and discussions.

\section{The matrix element of the electric current of a composite particle at the
lowest order in $\alpha $}
\indent
\par For simplicity, we assume that the lepton
is the bound state composed of a charged fermion and a neutral
scalar boson which we shall also call as preons for convenience.
We shall confine us in this article to the relativistic tightly
bounded states with $R^{-1}\gg m_l$ and $R^{-1}\geq m_F$, where
$R$ is the charge radius of the composite lepton and $m_F$ is the
mass of the charged fermionic preon. We shall consider two cases.
In the case A, the mass $m_F$ is of the same order as $R^{-1}$ and
much heavier than $m_l$. In the case B, $m_F$ and $m_l$  are both
much smaller than $R^{-1}$. In this case, if the force between the
preons is some confining  gauge interaction, there may be an
approximate global chiral symmetry which naturally results in the
small lepton mass $ m_l $. Essentially $R^{-1}$ is also the
confinement scale of the gauge interaction.

The general Lorentz and space inversion invariant B-S wave
function for a composite lepton composed of a fermionic and a
scalar preon is
\begin{eqnarray}\label{bs}
\chi^s_P(p)= \int d^4x e^{ip\cdot x} <0|T(
\psi(x/2),\phi(-x/2))|P,s>=(f_1+{i\hat{p}\over M}f_2)u^s(P),
\end{eqnarray}
where $\psi$ and $\phi$ are fields of the fermion and bosonic
preon respectively, $\hat{p}=p\cdot \gamma$, $P=p_1+p_2$ is the
momentum of the lepton, $p={m_Bp_1-m_Fp_2\over m_B+m_F}$ is the
relative momentum between preons, $m_B$ and $m_F$ are the masses
of the bosonic preon and fermionic preon respectively. The two
cases  mentioned above are: A. $m_F=O( R^{-1}) \gg m_l$. B.
$m_F,~m_l \ll R^{-1}$. In Eq. (\ref{bs}) the constant M with the
mass dimension is introduced for convenience, $f_i=f_i(P\cdot
p,p^2)\equiv f_i(P,p)$,  $(i=1,2)$, are real functions
corresponding to the S wave state, and $u^s(P)$ is the Dirac
spinor with spin component $s_z=s$.  It is straightforward to
derive from invariance under the space-time inversion that the BS
wave function for a lepton final state is
\begin{eqnarray}\label{be}
\bar\chi^s_P(p)=\bar u^s(P)(f_1+{i\hat{p}\over M}f_2).
\end{eqnarray}

The simplest diagram for the lowest order electro-magnetic
interaction of composite lepton is illustrated in Fig. 1. The
corresponding bound state matrix element is\cite{man,lur,pku,dh}
\begin{eqnarray}
\Gamma^{(0)}_\mu(P,q)&=&
<P+\frac{q}{2},s|J_{\mu}(0)|P-\frac{q}{2},s> \\ \nonumber
&=&-{1\over (2\pi)^4}\int d^4k\bar\chi_{P+q/2}(k+{q\over 4})
\cdot\gamma_\mu\chi_{P-q/2}(k-{q\over 4})i\Delta'^{-1}_F(({P\over
2}-k)^2) \\ \nonumber &=&-\bar u(P+q/2){1\over (2\pi)^4}\int
d^4k[f'_1+{i\over M}(\hat k+{\hat q\over 4})f'_2]
\\ \nonumber
&~&\times\gamma_\mu[f_1+{i\over M}(\hat k-{\hat q\over
4})f_2]i\Delta'^{-1}_F(({P\over 2}
-k)^2)u(P-q/2),\label{ma}
\end{eqnarray}
where
q is the momentum of the photon,
\begin{eqnarray*}
f_i=f_i(P-q/2, k-q/4),\; f'_i=f_i(P+q/2, k+q/4)\;\; for\; i=1,2,
\end{eqnarray*}
and $\Delta_F^{\prime}$ is the propagator of the scalar preon.

Using the Dirac equation, it is easy by a straightforward calculation to obtain
\begin{eqnarray}\label{vet}
\Gamma^{(0)}_\mu=\bar u(P+q/2)\{\gamma_\mu F^{(0)}_1(q^2)-{i\over
2}[\gamma_\mu,\hat q]
F^{(0)}_2(q^2)\}u(P-q/2),
\end{eqnarray}
where
\begin{eqnarray}\label{f1}
F^{(0)}_1(q^2)&=&S_1+{2T\over M^2}-{2m_l\over M}V_{1P}+{1\over
M^2}(m^2_l-{q^2\over 4})
T_{PP}+{q^2\over M^2}T_{qq}\\\nonumber
&&+{1\over 4}{q^2\over M^2}V_{2P}-{1\over 16}{q^2\over M^2}S_2,
\end{eqnarray}
\begin{eqnarray}\label{f2}
F^{(0)}_2(q^2)={m_l\over M^2}T_{PP}+{1\over 2M}S_3-{1\over
M}V_{1P}+{2\over M}V_{1q}
-{m_l\over 2M^2}V_{2P}.
\end{eqnarray}
In eqs. (\ref{f1}) and (\ref{f2}) the Lorentz invariant functions
$S_i$ etc. are defined by
\begin{eqnarray*}
S_1&=&{-i\over (2\pi)^4}\int d^4k f'_1f_1\Delta'^{-1}_F((P/2-k)^2),\\
S_2&=&{-i\over (2\pi)^4}\int d^4k f'_2f_2\Delta'^{-1}_F, \\
S_3&=&{-i\over (2\pi)^4}\int d^4k f'_2f_1\Delta'^{-1}_F,
\end{eqnarray*}
\begin{eqnarray} \label{sf}
{-i\over (2\pi)^4}\int d^4k
f'_2f_1\Delta'^{-1}_Fk_\mu=V_{1P}P_\mu+V_{1q}q_\mu,\\\nonumber
{-i\over (2\pi)^4}\int d^4k
f'_2f_2\Delta'^{-1}_Fk_\mu=V_{2P}P_\mu,\\\nonumber
{-i\over (2\pi)^4}\int d^4k f'_2f_2\Delta'^{-1}_Fk_\mu k_\nu=T_{PP}P_\mu
P_\nu+
T_{qq}q_\mu q_\nu+T \delta_{\mu\nu}.
\end{eqnarray}
The left hand side of the last two formulae in Eq. (\ref{sf}) is even in $q$,
hence there is no term linear in $q_\mu$ on the right hand side.

It follows from Eq. (\ref{f1}) that the normalization condition of
the electric charge is
\begin{eqnarray}
S_1(0)+{2\over M^2}T(0)-{2m_l\over M}V_{1P}(0)+{m^2_l\over
M^2}T_{PP}(0)=1,\label{nor}
\end{eqnarray}
which can approximately be written as
\begin{eqnarray}\label{norm}
S_1(0)+{2\over M^2}T(0)=1
\end{eqnarray}
if $m_l/M \ll 1$. In Eq. (\ref{norm})
\begin{eqnarray}\label{T0}
T(0)={1\over4} {-i\over (2\pi)^4}\int d^4k f_2^2\Delta'^{-1}_Fk^2.
\end{eqnarray}

Let us choose M to make the integral of $f_if_j\Delta'^{-1}_F$ to
be of the same order for $i,j=1,2$. From (\ref{T0}) we have
\begin{eqnarray}\label{TS}
T(0)=O(R^{-2}) S_1(0).
\end{eqnarray}
Since $V_{1p}(0)$ and $V_{1q}(0)$ are of the same order of
$S_1(0)$, which (as well as Eq. (\ref{TS})) has been checked by
using the spectra representation (\ref{spe}) given below, we
obtain from (8),(9),(12) and (14)
\begin{eqnarray}\label{F2}
{F_2(0)\over F_1(0)}={{C_2 M}\over {M^2+C_1R^{-2}}}
\end{eqnarray}
where $C_1$ and $C_2$ are constants of the order one.

Eq. (\ref{F2}) is in agreement with the result in Eq. (25) of
\cite{bd} derived with a different approach. Authors of \cite{bd}
asserted that the constant $M$ with the dimension of mass should
be equal to $m_F$. In our opinion, $M=m_F$ is likely to hold in
theories with vector or pseudo-vector interactions as a chirality
flip along the fermion line is required for the $F_2$ term.
However, it may not be true in theories with scalar or
pseudo-scalar interactions which  change the chirality. In the
latter case the more natural possibility is $M=O(R^{-1})$. Thus,
in general we have $M=m_F$ or $O(R^{-1})$. In case A, $m_F$ and
$R^{-1}$ are of the same order, from Eq. (15) the anomalous
magnetic moment of the muon arising from compositeness at the
leading order of $\alpha$ is

\begin{eqnarray}\label{a1}
a_\mu=O({m_l\over R^{-1}}).
\end{eqnarray}

In case B, $m_F\ll R^{-1}$, from Eq. (15) one has
\begin{eqnarray}\label{a2}
a_\mu=O({m_l\over R^{-1}}),
\end{eqnarray}
for $M=O(R^{-1})$, or
\begin{eqnarray}\label{a3}
a_\mu=O({m_lm_F\over R^{-2}})
\end{eqnarray}
for $M=O(m_F)$.

Therefore, $a_\mu$ can be suppressed linearly or quadratically by
$O(R^{-1})$ depending on whether $m_F=O(R^{-1})$ or $m_F\ll
R^{-1}$ and also on different internal dynamics.

 From (\ref{a1}), (\ref{a2}) and (\ref{a3}), the magnetic moment of the lepton bound
state  is found to approximately be the Dirac magnetic moment
$e/{2m_l}$ at the zeroth order of $\alpha$ provided that the
charge radius of the lepton is small enough.

By drawing lines corresponding to the ``very strong'' interactions
which bind preons into a lepton in Feynman diagrams one can obtain
diagrams more complex  than Fig. 1, an example for such diagrams
is shown in Fig. 3(a) . The contributions from such diagrams do
not change the estimate obtained above. We shall discuss such
diagrams in the next section.

\section{Radiative corrections at the $\alpha$ order  }
\subsection{The contribution corresponding to the simplest diagram}
\indent
\par It is necessary to examine the radiative corrections at higher orders of
$\alpha$ in the composite model in view of the agreement between
the experimental data and the standard model prediction for
$a_\mu$ up to the order $10^{-8}$ which is of the order $({\alpha
\over \pi})^3$. In this section, we will consider radiative
corrections to the anomalous magnetic moment of the composite
lepton at the $\alpha$ order.

To illustrate how the corrections at the $\alpha$ order in the
composite model is suppressed let us consider at first the case A
, $m_F=O(R^{-1})\gg m_l$. For simplicity we set $m_F=m_B=m$ in
this section. The simplest diagram at this order is shown in Fig.
2 and the corresponding matrix element is given by
\begin{eqnarray}\label{mat}
\Gamma_{\mu}^{(1)}= -{i\over (2\pi)^4}\int
d^4p\bar\chi_{P+q/2}(p+{q\over 4})\Lambda_\mu\chi_{P-q/2}
(p-{q\over 4})\cdot\Delta'^{-1}_F(({P\over 2}-p)^2),
\end{eqnarray}
where
\begin{eqnarray}
\Lambda_\mu={ie^2\over (2\pi)^4}\int d^4k {1\over k^2}\gamma_\nu{-i(\hat
p'_1-\hat k)
+m\over (p'_1-k)^2+m^2}\gamma_\mu{-i(\hat p_1-\hat k)+m\over
(p_1-k)^2+m^2}\gamma_\nu,\\
p_1={P\over 2}+p-{q\over 2},\;\; p'_1={P\over 2}+p+{q\over 2}.\nonumber
\end{eqnarray}

Firstly, let us consider the contribution of the $f_1$ term in the
BS wave function (\ref{bs}). Using both the Dirac equation and the
symmetry of integrand under permuting Feynman parameters,
 it is easy to show that (\ref{mat}) can be
transformed into the same form as Eq. (\ref{vet}), which is
expected from the conservation of electric current.

In calculating the anomalous magnetic moment $q^2$ can be put to
zero. The correction to $F_1(0)$ from Eq. (\ref{mat}) is absorbed
by the normalization condition $F_1(0)=1$. We know by examining
Eq. (\ref{mat}) that the contribution to $F_2(0)$ comes from the
term
\begin{eqnarray}\label{moma}
{\alpha\over \pi}m\bar u(P+q/2)u(P-q/2){-i\over (2\pi)^4}\int
d^4pf^2_1(P,p,q=0) \Delta'^{-1}_F(({P\over 2}-p)^2)\\\nonumber
\times \int_{x_1+x_2\leq 1}dx_1dx_2{2(1-x_1-x_2)({P\over 2}+p)_\mu
\over (x_1+x_2)[(1-x_1-x_2)({P\over 2}+p)^2+m^2]}.
\end{eqnarray}

>From the normalization condition (\ref{norm}) of the electric
charge and Eq. (\ref{TS}), we have
\begin{eqnarray}
S_1(0)=O(1)
\end{eqnarray}
because of $M=O(R^{-1})=O(m)$ in the case A.

Since $m=O(R^{-1}) \gg m_l$, $O(m_l/m)$ terms  can be neglected.
When the momentum of the internal preon ${P\over 2}+p$ is in the
Euclidean region,  the term in the square brackets in the
denominator of the integrand in the second line of (\ref{moma}) is
larger than $m^2$. Therefore, once the wave function can be
continued into the Euclidean region, as  verified by Wick
\cite{wick}, the contribution of (\ref{moma}) to the anomalous
magnetic moment is
\begin{eqnarray}\label{or}
O(\alpha{e\over 2m})=O(\alpha{e\over 2R^{-1}}).
\end{eqnarray}
The conclusions will not change when the contribution from the $f_2$ term of the
BS wave function (\ref{bs}) is included.

Similar analysis can be carried out for the case B, $m\ll R^{-1}$,
in which $m^2$ in the denominator of (\ref{moma}) can be neglected. The same result as Eq.
(\ref{or}) is obtained when $M=O(R^{-1})$.
Nevertheless, when $M=O(m)$, the
$({f_2\over M})^2$ term dominates and instead of (22) we have
\begin{eqnarray}\label{T1}
{T(0)\over M^2}=O(1).
\end{eqnarray}
The contribution of the $f_2^2$ term can be obtained by replacing
$f_1^2$ by $({f_2\over M})^2p^2$ in (\ref{moma}). Using (\ref{T1})
we find that the contribution of Fig.2 to the anomalous magnetic moment
is $O(\alpha{em\over 2R^{-2}})$.

\subsection{The contributions corresponding to complex diagrams}
\indent
\par In order to discuss more complex diagrams, we assume that the wave function
has the following spectrum representation\cite{wick}
\begin{eqnarray}\label{spe}
f_i(P,p)=\int\limits^\infty_{\mu^2_{min}}d\mu^2\int\limits^{{1\over2}}_{-{1\over
2}}
dy{g_i(y,\mu^2)\over [p^2-2yP\cdot p+\mu^2-i\epsilon]^{n_i}}, \;\; i=1,2,
\end{eqnarray}
where $\mu^2_{min}=m^2-{m^2_l\over 4}$. The advantage of this
representation is that the dependence of $f_i$ on the momenta $P$
and $p$ appear directly only in the propagator-like denominator.
In calculating diagrams containing B-S wave functions and
additional propagators one can use Feynman parameters to combine
all factors containing the momenta into a single factor and carry
out the integration over the internal momenta of the diagram. This
makes the dimensional estimate clear. For the purpose of
illustration we use the spectrum representation to the simplest
diagram, Fig. 2, discussed in last subsection. Substituting Eq.
(\ref{spe}) into Eq. (\ref{mat}) and using the approximation
\begin{eqnarray}
\Delta'^{-1}_F\simeq \Delta^{-1}_F=({1\over 2}P-p)^2+m^2,
\end{eqnarray}
the contribution of $f_1$  to the anomalous magnetic moment
becomes
\begin{eqnarray}\label{ano}
\alpha\int d\mu^2_1d\mu^2_2dy_1dy_2g_1(y_1,\mu^2_1)g_1(y_2,\mu^2_2)\int
\prod
\limits^5_{i=1}dx_i\delta(1-\sum\limits^5_{i=1}x_i)\\\nonumber
\times \{{mb_1(y_i,x_i)\over
[a(y_i,x_i)q^2+x_1\mu^2_1+x_2\mu^2_2+(x_3+x_4)m^2]^
{2n_1-2}}\\ \nonumber
+{m^3b_2(y_i,x_i)\over [a(y_i,x_i)q^2+x_1\mu^2_1+x_2\mu^2_2+(x_3+x_4)m^2]^
{2n_1-1}}\}
\end{eqnarray}
where $a$, $b_1$ and $b_2$ are rational functions of $x_i$ and
$y_i$ whose explicit expressions are omitted here.

Using the spectrum representation (\ref{spe}), it is easy to obtain
\begin{eqnarray}\label{s1}
S_1(q^2)&=&2^{-3}\pi^{-2}{(2n_1-4)!\over [(n_1-1)!]^2}\int
d\mu^2_1d\mu^2_2dy_1dy_2
g_1(y_1,\mu^2_1)g_1(y_2,\mu^2_2)\\\nonumber
&~&\times \int dx_1dx_2(x_1x_2)^{n_1-1}\delta(1-x_1-x_2)\\\nonumber
&~&\times\{{2n-3\over 2}{m^2\over [x_1x_2({1\over 2}-y_1)({1\over
2}-y_2)q^2+
x_1\mu^2_1+x_2\mu^2_2]}+1\}\\\nonumber
&~&\times {1\over[x_1x_2({1\over 2}-y_1)({1\over
2}-y_2)+x_1\mu^2_1+x_2\mu^2_2]^{2n_1-3}}.
\end{eqnarray}
Comparing
(\ref{ano}) with (\ref{s1}),
 the conclusion that the $\alpha$ order correction of the anomalous magnetic moment from
 the diagram in Fig. 2 is
given by  (\ref{or}) in the case A
 is obtained once again, as expected.

 Similar analysis can be carried
out for the case B, $m_F\ll O(R^{-1})$. The difference is that in
the case B the contribution of the $f_2^2$ term  dominates if
$M=m_F$.

When the "very strong" interaction binding preons into a lepton is
considered, it will bring corrections to the electro-magnetic
vertex of the lepton. Accordingly, there will appear much more
complex diagrams. If we assume that the constituent fermion
interacts with the constituent boson through exchanging bosons, we
get complex diagrams, for example, those in Fig. 3 at the lowest order of $\alpha$.
Such diagrams can be
calculated with the procedure similar to that used to obtain (23).
Combining the denominators of the integrand in the
electro-magnetic vertex with Feynman parameters and integrating
out the internal momentum, there always appear terms with a factor
$(x_1\mu^2_1+x_2\mu^2_2+\sum\limits^N_{i=3}x_im^2)^L$
 in the denominator in the approximation $\mu_i^2 \gg
m^2_l,q^2$  or $m^2 \gg m^2_l,q^2$ . The corresponding numerator
is a polynomial in $m_l$, $m$ and $\mu_i$ with the appropriate
dimension. Since the effective value of $\mu _i^2$ is of the order
$R^{-2}$, the order of magnitude of the denominator is controlled
by the larger one of $R^{-2}$ and $m^2$. Since $L$ is increased by
1 for one additional propagator and correspondingly the order of magnitude of
the polynomial in the nominator is increased by 2  from the
dimensional reason, the contributions from all these diagrams to
the anomalous magnetic moment are at most of the order of that
from the diagram with the least number of propagators which is
Fig.1 at the zeroth order of $\alpha$ and is Fig.2 at the first
order of $\alpha$. If the exchanged boson is a scalar, we assume
that the coupling constant $ \lambda $ of three scalar particles
is of $O(m)$. If the exchanged boson is a gauge boson, the
derivative coupling does not change the result of the dimensional
analysis here.

The above discussion can also be applied to the diagrams in which
a bosonic preon is converted into a pair of fermionic preons. Under the
assumption that all preons have the same mass m, which is used in the above
discussion for simplicity, the result is of the same form as that described above.

From the above discussions we conclude that the contribution to
$F_2(0)$ from any individual diagram, which contains BS wave
functions and additional  finite number of propagators of the
preons or the particles mediating the "very strong" interaction,
is suppressed by $O(\alpha R)$ or $O(\alpha m R^2)$ at the order
$\alpha$. However, can the QED result for $a_\mu$ at the $\alpha$
order approximately be obtained in the composite model and how
large is the deviation from the QED result? This question is
answered as follows.

Now we return to the general case. Adding infinite diagrams, we
will get the pole term contribution of the bound state (see Fig.
4). The four point Green function ${\cal K}$ in Fig. 4 is the
Fourier transform of $<0|T(\psi (x_1)\phi (x_2)
\bar\psi(y_1)\bar\phi(y_2))|0>$. It can be written as
\begin{eqnarray}\label{h}
{\cal K}(K;p',p)&=&{\chi_K(p)\bar\chi_K(p')\over
K^2+m^2_l}+\cdots\\\nonumber
&=&(f_1(K,p)+{if_2(K,p)\hat p\over M}){-i\hat K+m_l\over
K^2+m^2_l}\\\nonumber
&~&\times (f_1(K,p')+{if_2(K,p')\hat p'\over M})+\cdots.
\end{eqnarray}
The terms which have not been written explicitly in Eq. (\ref {h})
correspond to contributions from possible excited states of the
lepton or states of free preons.  In the case A of heavy preons,
these terms can be neglected when $K^2\ll m^2=
O(R^{-2})$. In case B of light
preons, these terms can be neglected when $K^2$ is much smaller
than the scale of confinement, which is essentially equivalent to
$K^2\ll R^{-2}$. (If a lepton has high excited states accessible
through the electro-magnetic transition, it is required that $K$
is much smaller than the masses of excited states which are of the
same order as $R^{-1}$ ). Therefore,  the contributions of these
omitted terms to the anomalous magnetic moment are of the order
$O(R^{2})$ or $O(m^{-2})$. The sub-diagram $T$ in Fig. 4
represents a five point vertex function with two fermionic preons,
two bosonic preons and the electro-magnetic current. Each
sub-diagram $T$ in Fig. 4 combining with two factors\
$f_i,\;f_j^{^{\prime }}$ which are associated with the two
adjacent ${\cal K}$ diagrams according to (\ref {h}) forms an
on-shell electro-magnetic vertex $\Gamma _\mu ^{(0)}$ at the
zeroth order in $ \alpha $. Since in (\ref{vet}),
$F_1^{(0)}\approx 1$ and $F_2^{(0)}$ can be neglected when the
photon momentum squared is much smaller than $R^{-2}$ , and the
contribution from the region of large momentum squared $k^2$ of
the virtual photon in Fig. 4 is highly suppressed, we get the
electro-magnetic vertex at the $\alpha$ order corresponding to
Fig.4 \cite{dh}
\begin{equation}\label{g1}
\Gamma^{(\alpha)}_\mu = {ie^2\over (2\pi)^4}\int \bar
u(P')\gamma_\nu{-i(\hat P'+\hat k)+m_l\over
(P'+k)^2+m^2_l}\gamma_\mu{-i(\hat P+\hat
k)+m_l\over(P+k)^2+m^2_l}\gamma_\nu{d^4k\over k^2} u(P)   + O(\alpha R).
\end{equation}
Eq. (\ref{g1}) is the same as the radiative correction at the $\alpha$
order of the QED vertex to a point particle when R is small
enough.

\section{Conclusions and discussions}
\indent
\par The above calculations show that the anomalous magnetic moment of a
composite lepton up to
$e^3$ order is
\begin{eqnarray}\label{mu1a}
\mu={e\over 2m_l}[{\alpha\over 2\pi}+O(\alpha^2)+O({m_l\over R^{-1}})]
\end{eqnarray}
in the case A and it is
\begin{eqnarray}\label{mu1}
\mu={e\over 2m_l}[{\alpha\over 2\pi}+O(\alpha^2)+O({m_l\over
R^{-1}})]
\end{eqnarray}
or
\begin{eqnarray}\label{mu2}
\mu={e\over 2m_l}[{\alpha\over 2\pi}+O(\alpha^2)+O({m_lm_F \over
R^{-2}})].
\end{eqnarray}
depending on the internal dynamics in the case B.

Apparently, these results may be generalized to arbitrary
order in $\alpha $ and we have
\begin{eqnarray}\label{aa}
a_\mu =a_\mu ^{SM}+O(\frac{m_l}{R^{-1}})
\end{eqnarray}
for the case A, and
\begin{eqnarray}\label{ab}
a_\mu =a_\mu ^{SM}+O(\frac{m_l} {R^{-1}})
\end{eqnarray}
for theories without a chiral symmetry or
\begin{eqnarray}\label{ac}
a_\mu =a_\mu ^{SM}+O(\frac{m_lm_F} {R^{-2}})
\end{eqnarray}
 for theories of the
confining gauge interactions with an approximate global chiral
symmetry in the case B. In Eq. (\ref{aa}), (\ref{ab}) and
(\ref{ac}) we have assumed $R^{-1}\gg m_Z$, which is necessary
when including the electro-weak corrections. In the recent
literature, e.g., the refs.~\cite{cm} and \cite{lane}, it is stated that the
corrections from compositeness to $a_{\mu}$ is of the order
$O({m_l^2 \over R^{-2}})$. This corresponds to the case (\ref{ac})
if $m_F$ is of the same order as $m_l$.

Similar conclusions can be obtained if the scalar preon is
charged. In contrast with the non-relativistic loose bound state ,
the magnetic moment of a  relativistic tight bound state
 is not of the vectorial sum of the magnetic moments of its
constituents. When the radius of the bound state is sufficiently
small, the bound state behaves as a whole in electro-magnetic
field, like a point particle does. Although our conclusions are
obtained in a simple composite model, they depend only on the
dimensional analysis and the analytical property of the BS wave
function. So it is expected that the conclusions will stand also
in other more complex models.

It is obvious from Eqs. (\ref{aa}), (\ref{ab}) and  (\ref{ac})
that the observed deviation of $a_{\mu}$ can be a demonstration of
the substructure of the muon and give a constraint on the radius
of the muon or the masses of preons.
 From the theoretical Eqs. (\ref{aa}), (\ref{ab}),
(\ref{ac}) and the experimental results,  we get
\begin{eqnarray}
R^{-1}\succeq 10^8 m_{\mu}
\end{eqnarray}
for theories without an approximate chiral symmetry, or
\begin{eqnarray}
R^{-1}\succeq 10^4m_{\mu}
\end{eqnarray}
for composite leptons bound by gauge interactions with an
approximate global chiral symmetry if $m_F$ is of the same order
as $m_l$.

Surely, when the E821 result is regarded as the consequence of the
substructure of leptons, it is necessary to re-analyze the
features of leptons, especially their decay characters, in
composite models and probe the fundamental theory which governs
the dynamic of composite models in the deeper layer of the
structure of matter. One has started the research in the direction ~\cite{calmet}.

\section*{Acknowledgments}
The work was partly supported by the national natural science
foundation of China.

\newpage
\begin{figure}[htb]
\begin{center}
\epsfig{file=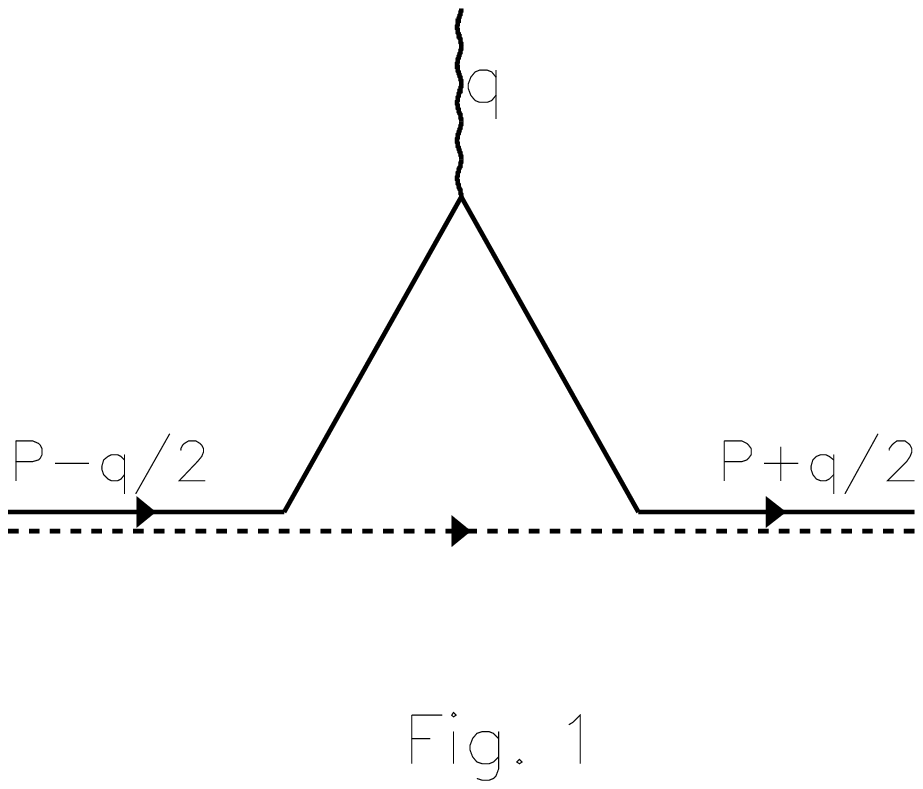,width=8cm}
\end{center}
\caption{ The lowest order diagram of the electro-magnetic interaction of
a composite lepton in the composite model of leptons. The solid (
dashed) line represents the fermionic (bosonic) preon and the wave line
represents the photon.
}
\end{figure}

\vspace{0.6 cm}

\begin{figure}[htb]
\begin{center}
\epsfig{file=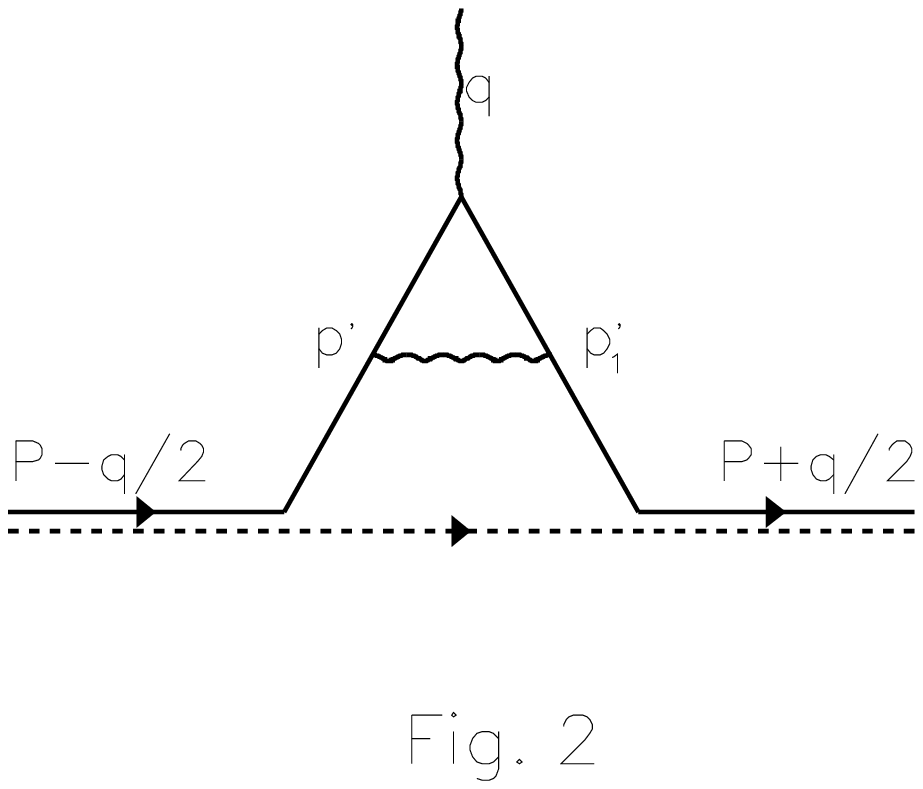,width=8cm}
\end{center}
\caption{ The $e^3$ order diagram of the electro-magnetic interaction of a
composite lepton in the composite model of leptons.
}
\end{figure}

\vspace{0.6 cm}

\begin{figure}[htb]
\begin{center}
\epsfig{file=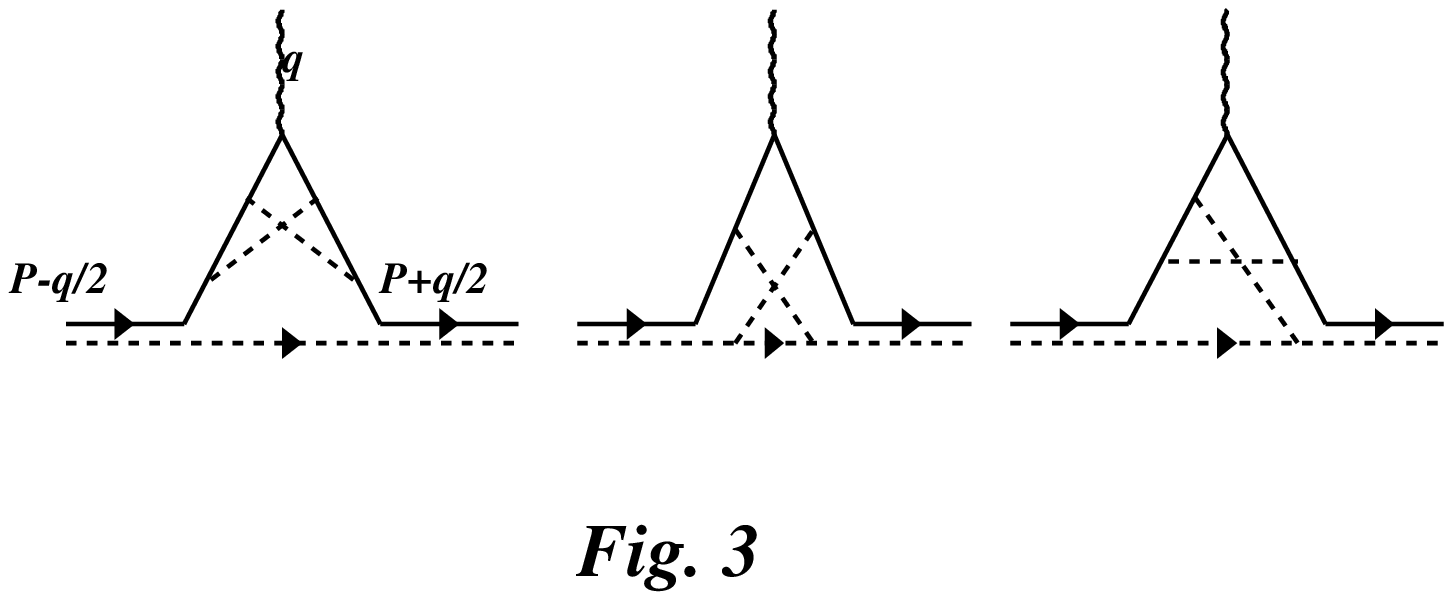,width=8cm}
\end{center}
\caption{ Some examples of complex diagrams. The dotted line
represents the boson which mediates the "very strong" interaction.
}
\end{figure}

\vspace{0.6 cm}
\begin{figure}[htb]
\begin{center}
\epsfig{file=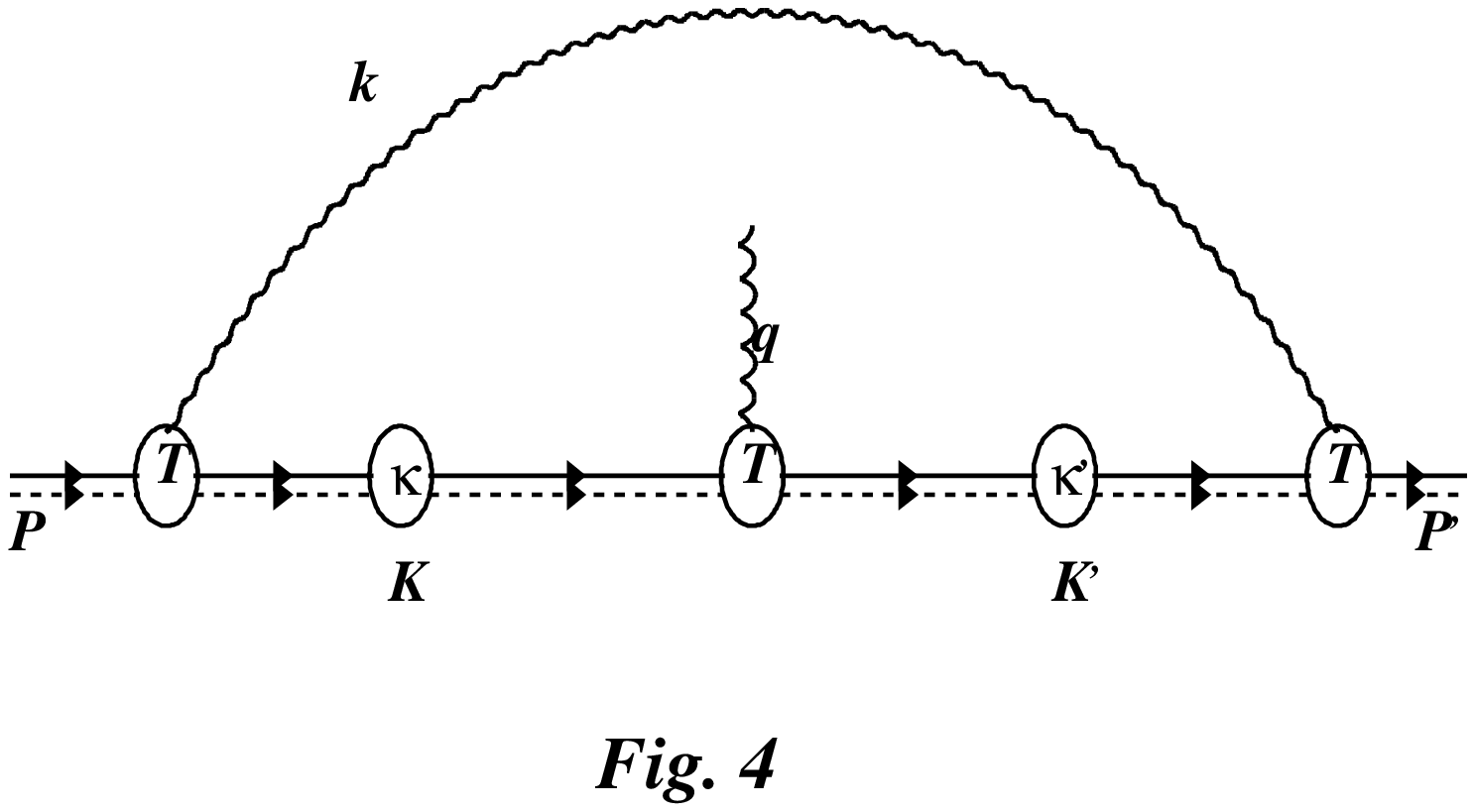,width=8cm}
\end{center}
\caption{ The diagram corresponding the contribution of the bound state
pole term. The circles with "T" and "${\cal K}$" (or "${\cal K'}$") inside
it
denote the electro-magnetic vertex of the bound state at the lowest order
in $\alpha$ when $K^2 (or K^{'2})=m_l^2$ and the 4-point Green function
respectively. K=P+k and K'=P'+k in the diagram
}
\end{figure}

\end{document}